# Oscillations of ultra-thin copper nanobridges at room temperature: Molecular dynamics simulations

Jeong Won Kang* and Ho Jung Hwang

Semiconductor Process and Device Laboratory, Department of Electronic Engineering,

Chung-Ang University, 221 HukSuk-Dong, DongJak-Ku, Seoul 156-756, Korea

We have investigated several ultra-thin copper nanobridges between supporting layers using a classical molecular dynamics simulation and a many-body potential function of the second-moment approximation of tight-binding scheme. This investigation has shown a part of the thermal properties of nanobridges, the tension in the nanobridges, and the resonance of the nanobridges. When the nanobridge has a well-defined structure, the resonant frequency is defined and the phenomenon of resonance is in common with classical oscillation systems.

PACS codes: 61.82.Rx, 02.70.Ns, 83.20.Jp

*E-mail: gardenriver@korea.com
Tel: 82-2-820-5296
Fax: 82-2-825-1584

# 1. I. Introduction

As the scale of microelectronic engineering continues to shrink, the fundamental interest has focused on the nature of the mechanical and electrical behavior of one-dimensional nanometer-scale channels such as quantum wires [1] and carbon nanotubes [2,3]. The development of new experimental techniques such as mechanically controllable break junctions (MCBJ) and the scanning tunneling microscope (STM) has made possible the formation and study of atomic-size junctions or contacts between macroscopic metals. The mechanical and electrical properties of nanocontacts between metallic bodies have been a subject of intensive research [4-13]. Quantum point contacts are structures in which a 'neck' of atoms just a few atomic diameters bridges two electrical contacts. The results prepared by contacting a metal surface with STM [7-14] and by other methods [15-19] typically have displayed a conductance quantized in steps of $2e^2/h$, where $e$ is the electron charge and $h$ is Planck's constant. The mechanical properties of nanocontacts have shown that before the first yielding, nanowires preserve the elastic stage, and after that, the elongation deformation proceeds in alternating quasi-elastic and yielding stages [12,13,20,21].

Nanobridges can be sustained by the supporting layers contacting with the both ends of nanobrigdes. This condition is considered as a similar situation in which a string, such as a guitar string, is connected between two clamps separated by a fixed distance. Therefore, the nanobridge can make a oscillation at a resonant frequency to set up a standing wave pattern. This paper investigates the oscillations of nanobridges using a classical molecular dynamics (MD) simulation, and presents the properties related to the resonance of the nanobridges.

## 2. Method

The MD simulations used the same MD methods in our previous works [20-26] with time step of 0.5 fs. The MD code used the velocity Verlet algorithm, a Gunstern-Berendsen thermostat to keep constant temperature, a periodic boundary condition along the wire axis, and neighbor lists to improve computing performance [27]. For Cu, we have used a many-body potential function of the second-moment approximation of tight-binding scheme [28] that has already been tested in nanoclusters [29-32] and nanowires [21,22,33], etc. The potential reproduces many basic properties of crystalline and noncrystalline bulk phases and surfaces [34], and gives a good insight into the structures and thermodynamics of metal clusters [35,36]. The total potential energy of the system, $E_{tot}$, can be expressed as a sum of the potential energies of each atom $i$. The energy of each atom, $E_i$, is expressed by the sum of the attractive band energy determined by the occupied local density of states, $E_i^{band}$, and the the repulsive term determined by the ion core–core interaction, $E_i^{rep}$. The total energy of system, $E_{tot}$, can be written as:

$$E_{tot} = \sum_i E_i = \sum_i (E_i^{rep} + E_i^{band}) \qquad (2\text{-}1)$$

$$E_i^{rep} = \sum_j A \exp[-p(r_{ij}/r_0 - 1)] \qquad (2\text{-}2)$$

$$E_i^{band} = \left[\sum_j \xi_0^2 \exp[-2q(r_{ij}/r_0 - 1)]\right]^{1/2} \qquad (2\text{-}3)$$

where $r_0$ and $r_{ij}$ are the nearest-neighbor bond length in a perfect crystal and the distance between $i$ and $j$ atoms, respectively. $A, \xi, p,$ and $q$ are adjustable parameters which are determined by fitting the calculated values of cohesive energy, lattice constant, and elastic constants to the experimental values. The parameters are same as those in Ref. [28]. The cut off distance is set to the value between the fourth and fifth nearest neighbors of perfect crystal.

The nanostructures we studied have two supporting layers that are connected with the both ends of the nanobridges, as shown in Fig. 1. The $D$ and $L$ in Fig. 1 are the diameter and the length of nanobridges, respectively. Last layers at both supporting ends are rigid; all the atoms in these layers are kept fixed during the MD simulations. These fixed layers at both ends are assumed to be connected to the external agent. Atoms in the following two layers adjacent to the fixed ones and those of the nanobridge are identified as dynamic atoms and are fully relaxed during the MD steps. We considered eight different structures, which are described in Table 1. The nanobridges with fcc structure have been observed in previous work and the ultra-thin nanowires also have cylindrical multi-shell (CMS)-type structure [22,33]. Therefore, we select the nanobridges with fcc or CMS-type structures. The supporting layers have the same structures as that of the nanobridges. Since the CMS-type nanobridges are similar to the {111} structure [22], they are connected with the supporting layers with {111} planes. Nanobridges with different diameter are simulated on condition of the same structure and same length.

### 3. Results and Discussion

The constant-temperature MD simulations have been performed during 1 ns at room temperature. In these simulations, the effect of thermal fluctuation is negligible. Figure 2 shows vibrations of the center of the nanobridges during initial 200 ps for all cases. The directions of vibrations shown in Fig. 2 are perpendicular to the axis of the nanobridges. In the cases of Al, A2, B1, B2, and C2, the vibrations of the nanobridges show the sinusoidal waves, such as resonance. In the A1 case, the resonance is stabilized after 100 ps. However, in the cases of C1, D1, and D2, the vibrations of the nanobridges disappear as the MD time increases. In classical oscillation systems, for certain frequencies, the interference produces a standing wave pattern or oscillation mode, and then the string resonates at the certain frequency, called resonant frequency. If the string is oscillated at some frequencies other than a resonant frequency, a standing wave is not set up but finally disappear. In our simulations of the nanobridges, if the nanobridges are in the mode of the certain resonant frequency in the beginning of the MD simulation, the standing waves of the nanobridges has been maintained. However, during the MD simulation, when the structure of the nanobridge was changed or atoms in the contact region between the supporting layer and the nanobridge were rearranged, the standing wave has not been set up in our simulations. As the diameter of the nanobridge increases, since the tension in the nanobridge is proportional with the diameter of the nanobridge, the displacement of its center decreases, and at the same times, the amplitude of wave decreases.

Figure 3 shows the structures after the MD simulations of 1 ns. In the A1 case, the neck of the nanobridge connected by the bottom-supporting layer became narrow, as indicated by the arrow. Therefore, the vibration before 100 ps has been damped as shown in Fig. 2. However, since its final structure is the same of the original structure, the resonance has been maintained in the nanobridge. In the C1 case, the nanobridge that originally has the {111} structure has been transformed into a complex structure with several connections. Especially, the region indicated by the arrow is a multi-shell-type 10-5-1 structure. Considering the 10-5-1 multi-shell nanoparticle, the pentagonal small particles, such as the decahedron and icosahedron particles, have been intensively investigated and reviewed in Ref. [37], and pentagonal multi-shell nanowires have been investigated quite recently [26,38]. The decahedron and icosahedron nanoclusters, called multiple twinned particles, are different from single crystalline particle and are composed of the tetrahedrons, which consist of only triangular {111} faces.

In the D1 and D2 cases, at the early stage of the MD simulations, the structural rearrangement has been shown in the contact regions between the supporting layer and the CMS-type nanobridge. As the results of the structural rearrangements, the deviations of the center of the nanobridges are shown in waves of Fig. 2. Most of all, the CMS-type nanobridges maintain their structures between the supporting layers with {111} planes. The lines in the D1 and D2 cases of Fig. 3 indicate the regions of CMS-type structure.

We also performed the MD simulations of 100 ps using the nanobridges, which have been obtained from simulations as explained in Fig. 2. The waves on the nanobirdge can have two modes such as longitudinal and transverse modes due to its structural property. The waves shown in Fig. 2 are closely related to the transverse wave on the nanobridges. Since the transverse wave on the nanobridge makes a resonant as shown in Fig. 2, the lowest resonant frequency of the nanobridge is calculated by following method. Atoms in the center regions of the nanobridges were initially applied to the external force, 0.01 nN, during 1 ps at room temperature. Then we monitored the displacements of the center region of the nanobirdge. Table 2 shows the lowest resonant frequency of the nanobridges at room temperature. In the C1 case, the resonance didn't appear. Therefore, it is assumed that the nanobridges, which are not a well-defined structure, cannot resonate at a certain frequency.

From the theory of classical oscillation system, the relation between wave length, $\lambda$, and $L$ for a standing wave on a string can be summarized as

$$\lambda = \frac{2L}{n}, \text{ for } n = 1,2,3,\cdots. \qquad (3\text{-}1)$$

The resonant frequencies that correspond to these wavelengths follow from Eq. (3-1):

$$f = \frac{v}{\lambda} = n\frac{v}{2L}, \text{ for } n = 1,2,3,\cdots, \qquad (3\text{-}2)$$

where $v$ is the speed of the traveling waves on the string. The lowest resonant frequency is $f = v/2L$, which corresponds to $n = 1$.

The speed of the waves on the string is also expressed as follows

$$v = \sqrt{\frac{\tau}{\mu}}, \qquad (3.3)$$

where $\tau$ is the tension in the string and $\mu$ is the linear density of the string. In the case of the first harmonic, the tension is calculated by Eqs. (3.2) and (3.3),

$$\tau = 4L^2 f^2 \mu. \tag{3.4}$$

In Table 2, the larger diameter of the fcc nanobridge is, the larger lowest-resonant-frequency is. When the wavelength is same, the speed of the traveling wave on the nanobridge increases in linearly proportional with the lowest resonant frequency. We can also calculate the linear density by using the mass of the nanobridge / the length of the nanobridge. From these values, the tensions in the nanobridges were calculated by Eq. (3.4) and are compared with those obtained from the MD simulations. As the diameter of the nanobridge increases, the tension in the nanobridge increases. The tensions obtained directly from the MD simulations are the forces acting on the supporting layers and are in agreement within the limits with those calculated by using the continuum theory, Eq. (3-4). For all cases, the tensions obtained from the MD simulations are slightly higher or lower than those obtained from the continuum theory. Since the tensions calculated by Eq. (3-4) reflect upon the only transverse waves, this difference could be derived from the interference of the longitudinal waves on the nanobridges. In spite of this difference, our results show that the mechanical properties of the nanobridges related to the oscillations of the nanobridges can be explained by the classical continuum theory of standing wave.

## 4. Conclusions

The nanobridges draw much interest for several reasons. One of the main reasons is that the nanobridge is the fundamental element to fabricate the nanometer-scale electronic devices. Another is the issue of their mechanical, electrical, and thermal properties. We have investigated a part of the thermal properties and the resonance of the nanobridges using classical molecular dynamics simulations. When the nanobridges have a well-defined structure, the resonance frequency is defined and the phenomenon of resonance is in common with classical oscillation systems. Although our study has disclosed some interesting features of nanobridges, further work including the effect of the longitudinal waves on the nanobridge has to be done.

Tables

Table 1. Structures of ultra-thin Cu nanobridges.

| | Supporting layer | Nanobridge | $D$ (Å) | $L$ (Å) | The number of atom in the nanobridge |
|---|---|---|---|---|---|
| A1 | {100} | {100} | 10.2 | 61.449 | 425 |
| A2 | {100} | {100} | 15.0 | 61.449 | 969 |
| B1 | {110} | {110} | 10.2 | 42.452 | 323 |
| B2 | {110} | {110} | 15.0 | 42.452 | 731 |
| C1 | {111} | {111} | 10.2 | 70.956 | 492 |
| C2 | {111} | {111} | 15.0 | 70.956 | 1 102 |
| D1 | {111} | CMS 6-1 | 4.9 | 64.294 | 209 |
| D2 | {111} | CMS 11-6-1 | 10.2 | 64.294 | 534 |

Table 2. The lowest resonant frequency, the speed of the traveling wave, the linear density, and the tension of the ultra-thin Cu nanobridges.

| Nanobridge | The lowest resonant frequency (GHz) | The speed of the traveling wave (m/s) | The linear density ($10^{-16}$ kg/m) | The tension in the nanobridge ($10^{-11}$ kg m/s$^2$) | |
|---|---|---|---|---|---|
| | | | | Eq. (3.4) | MD simulation |
| A1 | 5.6 | 68.9 | 73.0 | 3.46546 | 6.95621 |
| A2 | 7.2 | 89.2 | 166.5 | 13.24780 | 18.18869 |
| B1 | 15.6 | 131.0 | 80.3 | 13.78028 | 10.03054 |
| B2 | 18.5 | 155.4 | 118.8 | 28.68920 | 26.23471 |
| C1 | - | - | 73.2 | - | - |
| C2 | 8.3 | 117.8 | 163.9 | 22.75958 | 17.25665 |
| D1 | 4.7 | 60.2 | 34.3 | 1.24304 | 1.50018 |
| D2 | 7.8 | 99.8 | 87.7 | 8.73495 | 10.44815 |

Figures

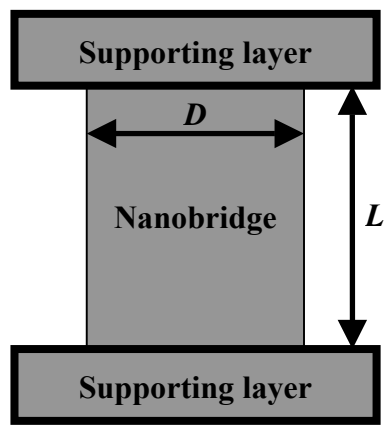

Figure 1. The Structure of nanobridge. The *D* and *L* are the diameter and the length of nanobridge, respectively.

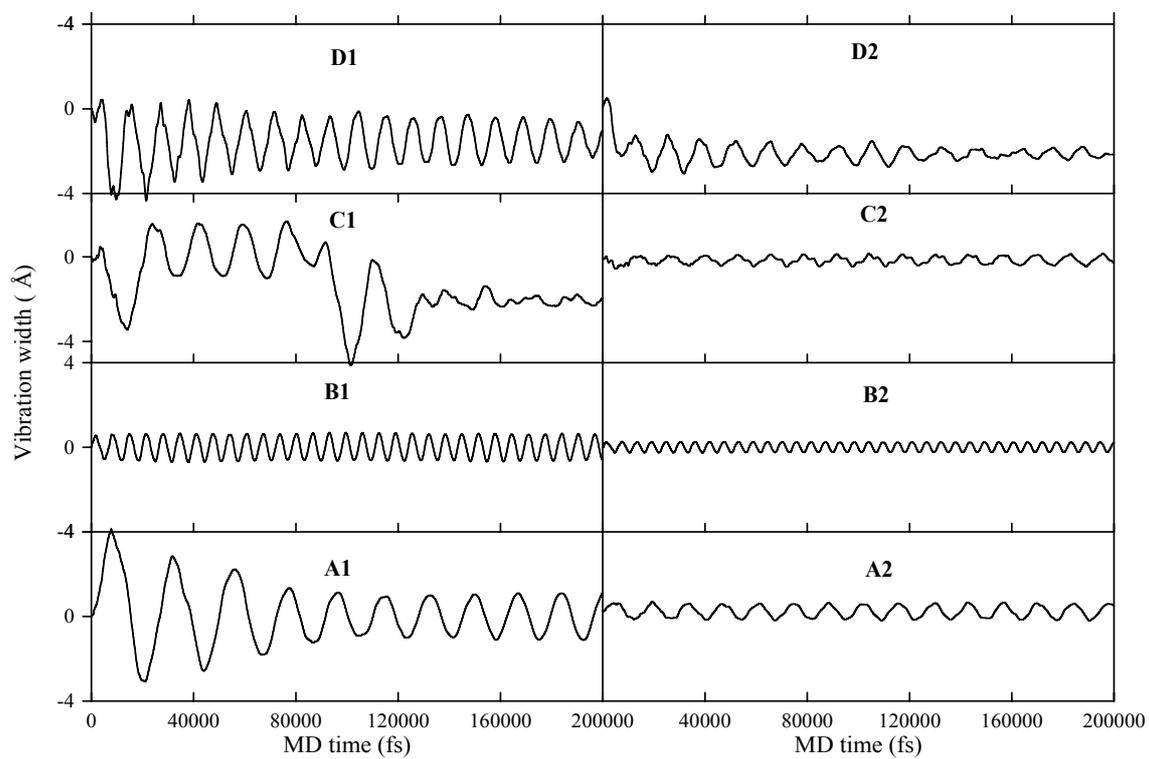

Figure 2. The vibration of the center of nanobridge during initial 200 ps for all cases in Table 1.

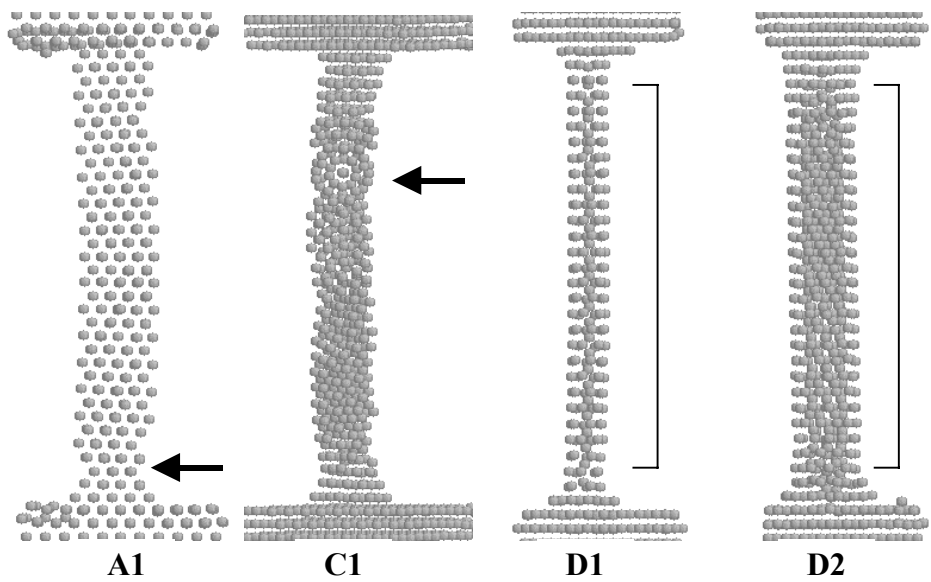

Figure 3. The structures after the MD simulations of 1 ns.